\documentstyle[aps,eqsecnum]{revtex}

\begin{document}

\title{Fast spin dynamics algorithms for classical spin systems}
\author{M. Krech, Alex Bunker, and D.P. Landau \\  Center for
Simulational Physics, University of Georgia, Athens, Georgia 30602}
\maketitle

\begin{abstract}
We have proposed new algorithms for the numerical integration of the
equations of motion for classical spin systems. In close analogy to
symplectic integrators for Hamiltonian equations of motion used in
Molecular Dynamics these algorithms are based on the Suzuki-Trotter
decomposition of exponential operators and unlike more commonly
used algorithms exactly conserve spin length and, in special cases,
energy. Using higher order decompositions we investigate integration
schemes of up to fourth order and compare them to a well established
fourth order predictor-corrector method. We demonstrate that these
methods can be used with much larger time steps than the
predictor-corrector method and thus may lead to a substantial
speedup of computer simulations of the dynamical behavior of magnetic
materials.

\end{abstract}
\draft


\section{Introduction}
Collective phenomena in many materials can be traced back to the
presence of interacting magnetic moments on the atomic level. The
exploration of magnetic systems therefore plays a key role in both
physics and materials science. The understanding of phase transitions,
critical phenomena, and scaling has in part been founded on the
investigation of simple model spin systems such as the Ising,
the XY, and the Heisenberg model (see below). These spin systems
continue to be of high relevance for the investigation
of dynamic critical behavior and dynamic scaling. On the other hand
realistic models of magnetic materials can be constructed from these
simple spin models, if the interaction parameters and
the underlying lattice structure are taken as an input from
experiments. However, in such cases the theoretical analysis of
experimentally accessible quantities, such as the dynamic structure
factor, is usually too demanding for analytical methods. Computer
simulations of the dynamical behavior of spin systems have, therefore,
become a very important tool for the theoretical understanding of
dynamic critical behavior and material properties of magnetic systems,
as the following examples may show.

Large scale computer simulations have been performed in recent years
in order to explore the dynamical behavior of classical XY
\cite{EvLan96} and Heisenberg models \cite{CosLan97} in $d = 2$
dimensions and of classical Heisenberg ferro- and antiferromagnets
in $d = 3$ \cite{ChenLan94,BunChenLan96}. The simulations are based on
model Hamiltonians for continuous degrees of freedom represented by a
three-component spin ${\bf S}_k$ with fixed length $|{\bf S}_k| = 1$
for each lattice site $k$. A typical model Hamiltonian is then given
by
\begin{equation}
\label{H}
{\cal H} = -J \sum_{<k,l>} \left(S_k^x S_l^x + S_k^y S_l^y + \lambda
S_k^z S_l^z \right) - D \sum_k \left( S_k^z \right)^2 ,
\end{equation}
where $J$ is the exchange integral, $<k,l>$ denotes a nearest-neighbor
pair of spins ${\bf S}_k$, $\lambda$ is an exchange anisotropy parameter, 
and $D$ determines the strength of a single-site or crystal field anisotropy.
For $\lambda = 1$ and $D = 0$ Eq.(\ref{H}) represents the classical
isotropic Heisenberg ferromagnet or the corresponding antiferromagnet
for $J > 0$ or $J < 0$, respectively. In the case $\lambda = D = 0$
Eq.(\ref{H}) reduces to the XY model.

Realistic descriptions of specific magnetic materials may require
additional interactions in the Hamiltonian, like an additional
two-spin exchange interaction between next nearest neighbors,
third nearest neighbors, etc. (see, e.g., Ref.\cite{Bohm80}). Two spin
interactions do not always provide a sufficient representation of the
interactions in a magnetic system. A more accurate description of the
isotropic ferromagnet EuS for example also requires a three spin
exchange interaction\cite{spin3} of the type
\begin{equation}
\label{H3}
{\cal H}_3 = - J_3 \sum_{<k,l,m>} \left({\bf S}_k \cdot {\bf S}_l
\right) \left({\bf S}_k \cdot {\bf S}_m \right) ,
\end{equation}
where $<k,l,m>$ denotes a triple of nearest neighbor spins. Four spin
exchange coupling constants are often negligible, but the
biquadratic interaction given by
\begin{equation}
\label{HBEG}
{\cal H}_4 = - J_4 \sum_{<k,l>} \left( {\bf S}_k \cdot {\bf S}_l
\right)^2 ,
\end{equation}
which is well known from the Blume - Emery - Griffith (BEG) model
\cite{BEG71}, needs to be considered in certain cases \cite{MFL96}.

The thermodynamic properties of the model can be obtained from a
Monte-Carlo simulation of the Hamiltonian ${\cal H}_{tot} = {\cal H} +
{\cal H}_3 + {\cal H}_4$ given by Eqs.(\ref{H}), (\ref{H3}), and
(\ref{HBEG}). In order to study the {\em dynamic} properties of the
spin system the equations of motion given by
\cite{EvLan96,CosLan97,ChenLan94,BunChenLan96}
\begin{equation}
\label{eqmot}
{d \over dt} {\bf S}_k = {\partial {\cal H}_{tot} \over \partial
{\bf S}_k} \times {\bf S}_k
\end{equation}
must be integrated numerically, where a Monte-Carlo simulation of the
model provides {\em equilibrium} configurations as initial conditions
for Eq.(\ref{eqmot}). Note that frequencies will be measured in energy
units so that $\hbar = 1$ in Eq.(\ref{eqmot}). A typical quantity to be
determined from the dynamics of the model is the dynamic structure factor
$S({\bf q},\omega)$, which is given by the space-time Fourier transform
of the spin-spin correlation function
\begin{equation}
\label{correl}
{\cal G}^{\alpha,\beta}({\bf r}_k - {\bf r}_l, t - t') \equiv \langle
S_k^{\alpha} (t) S_l^{\beta}(t') \rangle ,
\end{equation}
where $\alpha,\beta = x,y,z$ denote the spin component, ${\bf r}_k$
and ${\bf r}_l$ are lattice vectors, and the average $\langle \dots
\rangle$ must be taken over a sufficiently large number of independent
initial {\em equilibrium} configurations. The effect of collective
thermal excitations of the system (e.g., phonons) is not included in
the above approach. However, in magnetic systems the typical time scale
on which Eq.(\ref{eqmot}) has to be integrated is much shorter than
typical time scales set by these excitations, so that the above
procedure can be justified. In general one has to resort to the
solution of Langevin equations rather than Eq.(\ref{eqmot}) in order
to obtain the correct dynamics, but this is beyond the scope of this
article.

From the above outline of a typical Monte-Carlo spin dynamics study it
is evident, that at least in $d = 3$ the major part of the CPU time
needed for a spin dynamics simulation is consumed by the numerical
integration of Eq.(\ref{eqmot}). It is therefore desirable to do the
integration with the biggest possible time step. However, using
standard methods a severe restriction on the size of the time step is
posed by the accuracy within which the numerical method observes the
{\em conservation laws} of the dynamics. It is evident from
Eq.(\ref{eqmot}) that $|{\bf S}_k|$ for each lattice site $k$ and the
total energy are conserved. Symmetries of the Hamiltonian impose
additional conservation laws. If, for example, ${\cal H}_{tot} = {\cal
H}$ according to Eq.(\ref{H}) for $D = 0$ and $\lambda = 1$ (isotropic
Heisenberg model) the magnetization ${\bf M} = \sum_k {\bf S}_k$ is a
conserved quantity. For the anisotropic Heisenberg model, i.e.,
$\lambda \neq 1$ and/or $D \neq 0$ only the $z$-component $M_z$ of
the magnetization is conserved. Conservation of spin length and energy
is particularly crucial, because the condition $|{\bf S}_k| = 1$ is a
major part of the definition of the model and the energy of a
configuration determines its statistical weight. It would therefore
also be desirable to devise an algorithm which conserves these two
quantities {\em exactly}.

The outline for the remainder of this investigation is as follows. In
Sec.II we describe the properties of the currently used fourth-order
predictor-corrector method \cite{EvLan96,CosLan97,ChenLan94,BunChenLan96}.
Sec.III is devoted to the presentation of our new integration
procedure, based on Trotter-Suzuki decompositions of
exponential operators, which conserve spin length 
and, for the isotropic case only, energy
{\em exactly}. In Sec.IV we compare both schemes with special regard
to the accuracy within which the conservation laws hold and their
speed for a given demand on accuracy. A comparison
within the full framework of a spin dynamics simulation for a
Heisenberg ferromagnet is also presented. We finally discuss the
prospects of both algorithms in view of large scale spin dynamics
simulations in Sec.V.

\section{Predictor-corrector methods}
Predictor-corrector methods provide a very general tool for the
numerical integration of initial value problems like Eq.(\ref{eqmot})
with an equilibrium configuration as the initial value. The demand for
very good spin length and energy conservation, however, requires
methods with small truncation errors in the time step $\delta t$.
Further stability requirements for long times have led to the
implementation of a fourth-order scheme, which we briefly reproduce
here for the convenience of the reader.

In a more symbolic form Eq.(\ref{eqmot}) can be written as $\dot{y} =
f(y)$ with the initial condition $y(0) = y_0$, where $y$ is a
short-hand notation of a complete spin configuration written, e.g., as
a $3N$ dimensional vector for a lattice with $N$ sites. The initial
equilibrium configuration is simply denoted by $y_0$. The predictor
step of the scheme is then given, e.g., by the explicit Adams-Bashforth
four-step method \cite{NumAna81}
\begin{equation}
\label{AdBash}
y(t+\delta t) = y(t) + {\delta t \over 24} \left[55 f(y(t)) - 59
f(y(t-\delta t)) + 37 f(y(t-2\delta t)) - 9 f(y(t-3\delta t)) \right]
\end{equation}
which has a local truncation error of the order $(\delta t)^5$. The
corrector step consists of typically one iteration of the implicit
Adams-Moulton three-step method \cite{NumAna81}
\begin{equation}
\label{AdMoul}
y(t+\delta t) = y(t) + {\delta t \over 24} \left[9 f(y(t+\delta t)) +
19 f(y(t)) - 5 f(y(t-\delta t)) + f(y(t-2\delta t)) \right]
\end{equation}
which also has a local truncation error of the order $(\delta t)^5$.
The first application of Eq.(\ref{AdBash}) obviously requires the
knowledge of $y(\delta t)$, $y(2\delta t)$, and $y(3\delta t)$ apart
from $y(0) = y_0$. These can be provided by three successive
integrations of $\dot{y} = f(y)$ (see Eq.(\ref{eqmot})) by the
fourth-order Runge-Kutta method \cite{NumAna81} which cannot be used
for the entire time integration, because truncation errors accumulate
too fast during typical integration times required for high frequency
resolution. In order to apply Eqs.(\ref{AdBash}) and (\ref{AdMoul})
repeatedly the spin configuration at the last four time steps must be
kept in memory.

It is interesting to note that the predictor-corrector method according to
Eqs.(\ref{AdBash}) and (\ref{AdMoul}) observes the conservation laws for
the magnetization according to the symmetry of the Hamiltonian to within
machine accuracy if periodic boundary conditions are employed. The reason
is that all spins are updated simultaneously by Eqs.(\ref{AdBash}) and
Eq.(\ref{AdMoul}) and that the driving forces (torques) are evaluated
precisely according to the exact equation of motion (see Eq.(\ref{eqmot}))
for every time step. The periodic boundary conditions restore the discrete
translational invariance of the infinite lattice in the finite sample
needed for the simulation.

The predictor-corrector method is very general and, therefore, its
implementation is independent of the special structure of the right-hand
side of the equation of motion. Apart from the conservation of the
magnetization the other conservation laws discussed in
Sec.I will therefore only be observed within the accuracy set by the
truncation error of the method. In practice, this limits the time step
to typically $\delta t = 0.01/J$ in $d = 3$ \cite{BunChenLan96} for
the isotropic model Hamiltonian given by Eq.(\ref{H}) $(D = 0)$, where
the total integration time is of the order $200/J$. The same method
has also been used in $d = 2$ with a time step $\delta t = 0.025/J$
\cite{CosLan97}, but in this case the total integration time did not
exceed $60/J$. In view of the fact that a time resolution of typically
$0.1/J - 0.2/J$ is desired for the evaluation of the time
displaced correlation function given by Eq.(\ref{correl}), it is clear
that the total CPU time required for a spin dynamics simulation in $d
= 3$ is dominated by the numerical integration of the equations of
motion. A quantitative analysis is presented in Sec.IV.

\section{Suzuki-Trotter decomposition methods}
The motion of a spin according to Eq.(\ref{eqmot}) may be visualized
as a Larmor precession of the spin ${\bf S}$ around an effective axis
$\Omega$ which is itself time dependent. In order to illustrate our
reasoning we restrict ourselves again to the Hamiltonian ${\cal H}_{tot}
= {\cal H}$ given by Eq.(\ref{H}) for the simple case $D = 0$ but for
arbitrary values of $\lambda$. The evaluation of the right-hand side of
Eq.(\ref{eqmot}) then shows that the lattice can be decomposed into two
sublattices such that a spin on one sublattice performs a Larmor
precession in a local field $\Omega$ of neighbor spins which are {\em all}
located on the other sublattice. For the Hamiltonian given by Eq.(\ref{H})
there are only two such sublattices if the underlying lattice is, e.g.,
simple cubic or bcc.

The first basic idea of the algorithm to be described below is to
actually perform a {\em rotation} of a spin about its local field
$\Omega$ by an angle $\alpha = |\Omega|\delta t$, rather than
integrating Eq.(\ref{eqmot}) by some standard method. This procedure
guarantees the conservation of the spin length $|{\bf S}|$ to within
machine accuracy. (A procedure like this has already been implemented
in a numerical study of spin motions in a classical ferromagnet
\cite{WatBluVin69}). The second basic idea is to exploit the sublattice
decomposition of Eq.(\ref{eqmot}) to also ensure {\em energy}
conservation to within machine accuracy. Denoting the two sublattices
by $\cal A$ and $\cal B$, respectively, we can write Eq.(\ref{eqmot}) 
in the form
\begin{equation}
\label{eqmAB}
{d \over dt} {\bf S}_{k \in {\cal A}} =
\Omega_{\cal B}[\{{\bf S}\}] \times {\bf S}_{k \in {\cal A}}
\quad , \quad
{d \over dt} {\bf S}_{k \in {\cal B}} =
\Omega_{\cal A}[\{{\bf S}\}] \times {\bf S}_{k \in {\cal B}} ,
\end{equation}
where $\Omega_{\cal A}[\{{\bf S}\}]$ and $\Omega_{\cal
B}[\{{\bf S}\}]$ denote the local fields produced by the
spins on sublattice $\cal A$ and $\cal B$, respectively. Either of the
equations in Eq.(\ref{eqmAB}) reduces to a {\em linear} system of
differential equations if the spins on the other sublattice are kept
fixed. This suggests an {\em alternating} update scheme, i.e., the
spins ${\bf S}_{k \in {\cal A}}$ are rotated for the given values of
${\bf S}_{k \in {\cal B}}$ and vice versa. This implies that the
scalar products ${\bf S}_{k \in {\cal A}} \cdot \Omega_{\cal B}[\{{\bf
S}\}]$ remain constant during the update of ${\bf S}_{k \in {\cal A}}$
and the scalar products ${\bf S}_{k \in {\cal B}} \cdot \Omega_{\cal
A}[\{{\bf S}\}]$ remain constant during the update of ${\bf S}_{k \in
{\cal B}}$. From Eq.(\ref{H}) for $D = 0$ and Eq.(\ref{eqmot}) we
therefore find that the energy is {\em exactly} conserved during this
alternating update scheme (see also Eq.(\ref{Sktdt})). Note, that each
sublattice rotation is performed with the {\em actual} values of the
spins on the other sublattice, so that only a {\em single} copy of the
spin configuration is kept in memory at any time. However, the
magnetization will not be conserved during the above rotation
operations. Moreover, the two alternating rotation operations do not
commute, so that a closer examination of the sublattice decomposition
of the spin rotation is required.

In order to obtain a simple description of the operations performed on
the spin configuration during the integration of the equations of
motion we again represent a full configuration by a vector $y$ which
is decomposed into two sublattice components $y_{\cal A}$ and $y_{\cal
B}$ according to $y = (y_{\cal A},y_{\cal B})$. The cross products in
Eq.(\ref{eqmAB}) can be expressed by matrices $A$ and $B$ which are the
infinitesimal generators of the rotation of the spin configuration
$y_{\cal A}$ on sublattice $\cal A$ at fixed $y_{\cal B}$ and of the spin
configuration $y_{\cal B}$ on sublattice $\cal B$ at fixed $y_{\cal
A}$, respectively. The exact update of the configuration $y$ from time
$t$ to $t + \delta t$ can then be expressed by an exponential (matrix)
operator according to
\begin{equation}
\label{eAB}
y(t+\delta t) = e^{(A + B)\delta t} y(t) .
\end{equation}
Although the exponential operator in Eq.(\ref{eAB}) rotates each spin
of the configuration it has no simple explicit form, because the
rotation axis for each spin depends on the configuration itself (see
Eq.(\ref{eqmAB})) and is therefore not known {\it a priori}. However, the
operators $e^{A \delta t}$ and $e^{B \delta t}$ which rotate $y_{\cal
A}$ at fixed $y_{\cal B}$ and $y_{\cal B}$ at fixed $y_{\cal A}$,
respectively, {\em do} have a simple explicit form. We demonstrate
this for the case $\lambda = 1$ and $D = 0$ in Eq.(\ref{H}). For each
$k \in {\cal A}$ we find
\begin{equation}
\label{OmegaA}
\Omega_{\cal A}[\{{\bf S}\}] = -J \sum_{l = NN(k)} {\bf S}_l \equiv
\Omega_k ,
\end{equation}
where $NN(k)$ denotes the nearest neighbors of $k$ (which belong to
$y_{\cal B}$). Eq.(\ref{OmegaA}) can be readily generalized for
$\lambda \neq 1$, the case $D \neq 0$ will be discussed below. From
Eqs.(\ref{eqmAB}) and (\ref{OmegaA}) we obtain (see also
Ref.\cite{WatBluVin69})
\begin{equation}
\label{Sktdt}
{\bf S}_k(t+\delta t) = {\Omega_k (\Omega_k \cdot {\bf S}_k(t)) \over
\Omega_k^2} + \left[ {\bf S}_k(t) - {\Omega_k (\Omega_k \cdot {\bf
S}_k(t)) \over \Omega_k^2} \right] \cos (|\Omega_k|\delta t) +
{\Omega_k \times {\bf S}_k(t) \over |\Omega_k|} \sin (|\Omega_k|\delta t).
\end{equation}
Note, that according to Eq.(\ref{Sktdt}) $\Omega_k \cdot {\bf S}_k(t +
\delta t) = \Omega_k \cdot {\bf S}_k(t)$ which explicitly confirms
energy conservation. For $k \in {\cal B}$ Eq.(\ref{Sktdt}) also holds
in exactly the same form. The alternating update scheme for the
integration of Eq.(\ref{eqmAB}), i.e., Eq.(\ref{eqmot}) now
amounts to the replacement $e^{(A + B) \delta t} \to e^{A \delta t}
e^{B \delta t}$ in Eq.(\ref{eAB}), which is only correct up to terms
of the order $(\delta t)^2$ \cite{SuzUme93}. The magnetization will
therefore only be conserved up to terms of the order $\delta t$ (global
truncation error), which is insufficient for practical purposes. The
remedy for this shortcoming, which constitutes the third basic idea of
the algorithm, is now obvious. We simply employ higher order
Suzuki-Trotter decompositions of the exponential operator in
Eq.(\ref{eAB}) to increase the order of the local truncation error of
the algorithm and thus improve the magnetization conservation. The
simplest possible improvement is given by the well known second order
decomposition
\cite{SuzUme93}
\begin{equation}
\label{eA2BA2}
e^{(A + B) \delta t} = e^{A \delta t / 2} e^{B \delta t} e^{A \delta t
/ 2} + {\cal O}(\delta t^3),
\end{equation}
which will be used for comparison with the predictor-corrector method
outlined in Sec.II. Note, that Eq.(\ref{eA2BA2}) is equivalent to the
midpoint integration method applied to Eq.(\ref{eqmAB}) (see also
Ref.\cite{WatBluVin69}). We furthermore use the fourth order
decomposition \cite{SuzUme93}
\begin{equation}
\label{epABA}
e^{(A + B) \delta t} = \prod_{i=1}^5 e^{p_i A \delta t / 2} e^{p_i B
\delta t} e^{p_i A \delta t / 2} + {\cal O}(\delta t^5)
\end{equation}
with the parameters
\begin{equation}
\label{p5}
p_1 = p_2 = p_4 = p_5 \equiv p = 1/(4 - 4^{1/3}) \quad \mbox{and}
\quad p_3 = 1 - 4p.
\end{equation}
It is also possible to construct a decomposition like Eq.(\ref{epABA})
with only three factors, but then $|p_i| > 1$ for $i=1,2,3$ which is
numerically unfavorable \cite{SuzUme93}. From Eqs.(\ref{eA2BA2}) and
(\ref{epABA}) the close analogy to symplectic integrators obtained from
the Liouville operator formalism for Molecular Dynamics simulations
is obvious \cite{KPD97}. The sublattice decomposition of the spin degrees
of freedom in spin dynamics corresponds to the natural decomposition of
the degrees of freedom in Hamiltonian dynamics into positions and momenta.
Especially the second order decomposition given by Eq.(\ref{eA2BA2}) is
equivalent to the velocity Verlet algorithm for Molecular Dynamics
simulations \cite{V67} (see also Ref.\cite{KPD97} for recent developments
in this field) which itself is equivalent to the well-known leapfrog
algorithm \cite{KPD97}.

As will be shown in Sec.IV, the additional computational effort to be
invested in the evaluation of Eq.(\ref{epABA}) as compared to
Eq.(\ref{eA2BA2}) or Eqs.(\ref{AdBash}) and (\ref{AdMoul}) can be
compensated by using larger time steps. The evaluation of the
trigonometric functions in Eq.(\ref{Sktdt}) can also be avoided by
observing that the above decompositions are only correct to within a
certain order in $\delta t$. It is therefore sufficient to replace
$\sin x$ by its Taylor polynomial $T(x)$, where $T(x) = x$ if
Eq.(\ref{eA2BA2}) is used and $T(x) = x - x^3/6$ if Eq.(\ref{epABA})
is used. The cosine in Eq.(\ref{Sktdt}) then has to be replaced by
$\sqrt{1 - T^2(x)}$ in order to maintain conservation of spin length.
It is also worth noting that the above decompositions maintain the
time inversion property of $e^{(A + B) \delta t}$, i.e., their inverse
is obtained by the replacement $\delta t \to - \delta t$. A Hamiltonian
with both nearest and next-nearest neighbor two-spin interactions on,
e.g., simple cubic or bcc lattices, can be treated within the above
framework if the lattice is decomposed into four sublattices. In contrast
to the predictor - corrector method conservation of magnetization according
to the symmetry of the Hamiltonian is only observed within the truncation
error of the decomposition method, because according to Eqs.(\ref{eA2BA2})
and (\ref{epABA}) the sublattices ${\cal A}$ and ${\cal B}$ are no longer
strictly equivalent.

We now generalize the above considerations to the case $D \neq 0$ in
Eq.(\ref{H}). For a spin in sublattice ${\cal A}$ the equation of motion
reads
\begin{equation}
\label{eqmDAB}
{d \over dt} {\bf S}_{k \in {\cal A}} =
\Omega_{\cal B}[\{{\bf S}\}] \times {\bf S}_{k \in {\cal A}}
-2D S_{k \in {\cal A}}^z {\bf e}_z \times {\bf S}_{k \in {\cal A}},
\end{equation}
where ${\bf e}_z$ denotes the unit vector pointing along the $z$-axis,
and spins in sublattice ${\cal B}$ obey an equation of the same form. In
contrast to the isotropic case (see Eq.(\ref{eqmAB})) the equation of
motion for each individual spin on each sublattice is {\em nonlinear}.
It is possible to obtain an analytic solution of this equation, in the
spirit of Eq.(\ref{Sktdt}), where the trigonometric functions are basically
replaced by Jacobian elliptic functions, and to implement this solution as
a rotation operation in one of the above decomposition schemes. This
approach, however, involves very precision sensitive operations
which greatly limit speed and therefore the overall efficiency of such an
algorithm. In practice, it is much more convenient to include the
effects of the nonlinearity in Eq.(\ref{eqmDAB}) by an iterative
method, as described in the following paragraph.

For the sublattice decomposition of the spin rotation in the isotropic
case discussed above the requirement for energy conservation in the
presence of a single site anisotropy reads
\begin{equation}
\label{energyD}
\Omega_k \cdot {\bf S}_k(t+\delta t) - D \left[S_k^z(t+\delta t)\right]^2
 = \Omega_k \cdot {\bf S}_k(t) - D \left[S_k^z(t)\right]^2
\end{equation}
for $k \in {\cal A}$ and $k \in {\cal B}$, where $\Omega_k$ is given
by Eq.(\ref{OmegaA}). In order to perform a rotation operation in
analogy to Eq.(\ref{Sktdt}) we have to identify an effective rotation
axis. This can be achieved by rewriting Eq.(\ref{energyD}) in the form
$\widetilde{\Omega}_k \cdot ({\bf S}_k(t + \delta t) - {\bf S}_k(t)) =
0$, where $\widetilde{\Omega}_k$ is given by
\begin{equation}
\label{OmegaD}
\widetilde{\Omega}_k = \Omega_k - D \left(0,0,S_k^z(t) +
S_k^z(t+\delta t) \right),
\end{equation}
i.e., in order to perform the rotation $S_k^z$ at the future time $t +
\delta t$ must be known in {\em advance}. This problem can only be
solved iteratively starting from the initial value $S_k^z(t+\delta t)
= S_k^z(t)$ in Eq.(\ref{OmegaD}) and performing several updates according
to the decompositions given by Eqs.(\ref{eA2BA2}) or (\ref{epABA}),
respectively, in order to improve energy conservation according to
Eq.(\ref{energyD}). Naturally, this procedure leads to a substantial
slowdown of the integration algorithm, where the energy is no longer
exactly conserved. Details are discussed in Sec.IV. The biquadratic
interaction given by Eq.(\ref{HBEG}) can be treated by the same
iterative scheme. The three-spin interaction given by Eq.(\ref{H3}),
however, requires a reconsideration of the sublattice decomposition.
Specific numerical examples in comparison with those obtained from the
predictor-corrector method are presented in the following section.

\section{The methods in comparison}
For a quantitative analysis of the integration methods outlined
above we restrict ourselves to the Hamiltonian ${\cal H}_{tot} = {\cal
H}$ given by Eq.(\ref{H}) for $\lambda = 1$ in $d = 3$. The underlying
lattice is simple cubic with $L = 10$ lattice sites in each direction
and periodic boundary conditions are imposed in all cases discussed below. 
The simulations have been performed on IBM RS/6000 workstations at the
Center for Simulational Physics and a Cray C90 at the Pittsburgh
Supercomputer Center.

In order to compare the different integration methods we first
investigate the accuracy within which the conservation laws are
fulfilled. As a standard initial configuration we choose a well
equilibrated configuration from a Monte-Carlo simulation of the model
defined by Eq.(\ref{H}) for $\lambda = 1$ at a temperature $T = 0.8
T_c$ for $D=0$ and $D=J$, where $T_c$ refers to the critical
temperature of the isotropic model $(D = 0)$ and is given by
$K_c = J / (k_B T_c) = 0.693035$ \cite{KCL}. The magnetization of
such a configuration differs from zero and provides an indicator for
the numerical quality of the magnetization conservation. We integrate
the equations of motion to $t = 800/J$ and monitor the energy $e(t)
\equiv E(t)/(J L^3)$ of the configuration per spin and in units of the
exchange coupling constant $J$ and the modulus $m(t) \equiv
|{\bf M}(t)| / L^3$ of the magnetization per spin for 
the isotropic case $D = 0$ and its $z$-component $m_z(t) \equiv M_z(t)
/ L^3$ for the stongly anisotropic case $D = J$ as functions of time.
Note, that for these tests both integration methods are started from
identical initial configurations. The time step for the
predictor-corrector method is chosen as $\delta t = 0.01/J$ in all
cases.

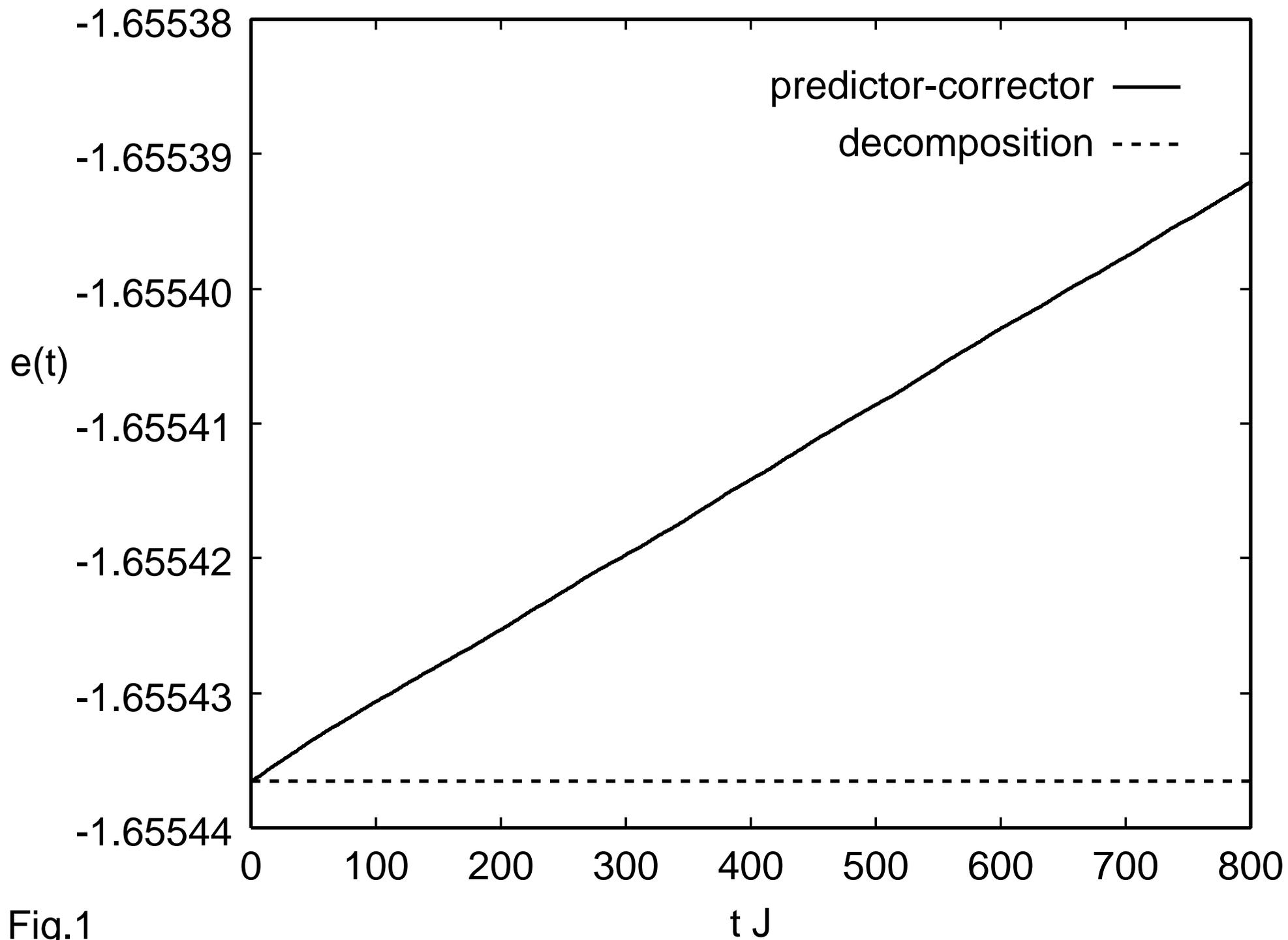
\begin{figure}[t]
\caption{
Energy $e(t) = E(t)/(J L^3)$ per spin for the predictor-corrector
method (see Eqs.(\protect\ref{AdBash}) and (\protect\ref{AdMoul})) for
the time step $\delta t = 0.01/J$ and $D = 0$ (solid line). Both
decomposition schemes (see Eqs.(\protect\ref{eA2BA2}) and
(\protect\ref{epABA})) conserve the energy exactly (dashed line).
\label{E2}}
\end{figure}

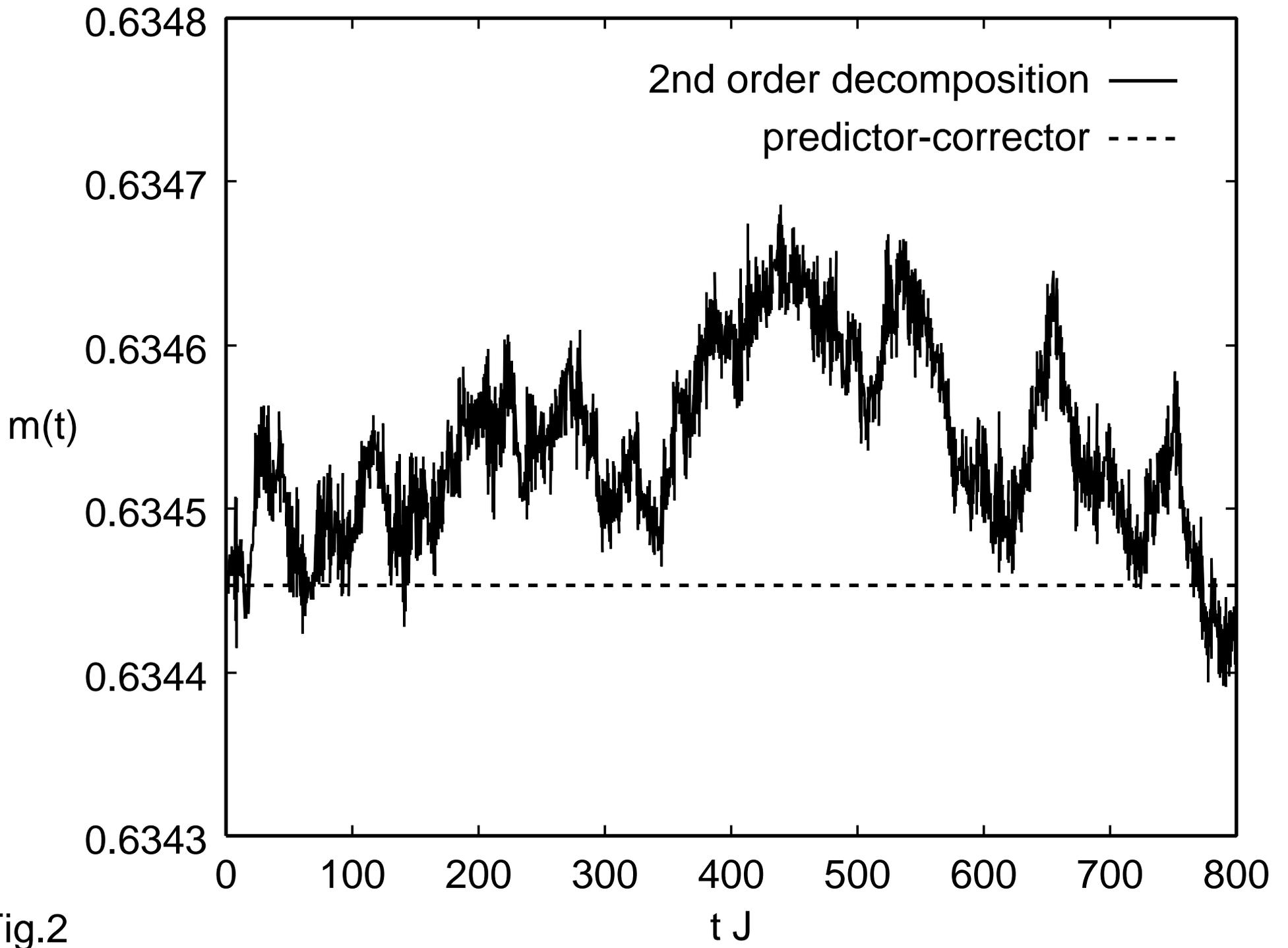
\begin{figure}[t]
\caption{
Magnetization $m(t) = |{\bf M}(t)|/L^3$ per spin for the
predictor-corrector method (see Eqs.(\protect\ref{AdBash}) and
(\protect\ref{AdMoul})) for the time step $\delta t = 0.01/J$ (dashed
line) and the second order decomposition scheme (see
Eq.(\protect\ref{eA2BA2})) for the time step $\delta t = 0.04/J$
(solid line) for $D = 0$. The predictor-corrector method conserves the
magnetization exactly, whereas Eq.(\protect\ref{eA2BA2}) yields
fluctuations in $m(t)$ on all time scales.
\label{Mz2}}
\end{figure}

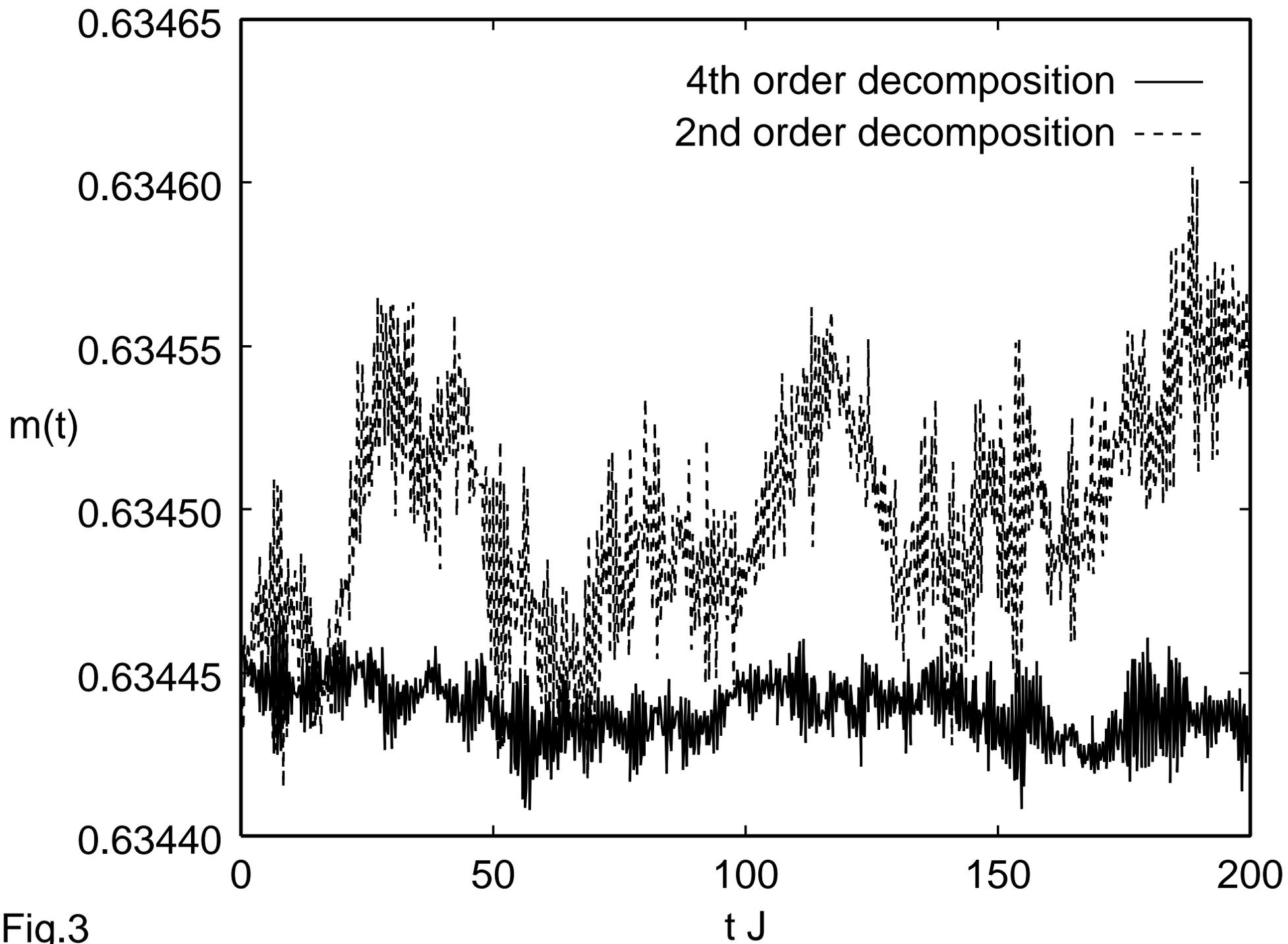
\begin{figure}[t]
\caption{
Magnetization $m(t)$ per spin for the fourth order decomposition
method (see Eq.(\protect\ref{epABA})) with $\delta t = 0.2/J$ (solid
line) and for the second order decomposition method (see
Eq.(\protect\ref{eA2BA2})) with $\delta t = 0.04/J$ (dashed line) for
$D = 0$. The CPU time needed for both methods with these time steps is
almost the same. Despite the large time step the fourth order
decomposition method yields substantially smaller magnetization
fluctuations than the second order decomposition.
\label{Mz4}}
\end{figure}

We first consider the case of $D = 0$ for which the implementation of
Eqs.(\ref{eA2BA2}) and (\ref{epABA}) using Eq.(\ref{Sktdt}) is
straightforward. The Taylor polynomial $T(x)$ is chosen as
$T(x) = x - x^3/6$ for Eq.(\ref{eA2BA2}) and $T(x) = x - x^3/6 +
x^5/120$ for Eq.(\ref{epABA}) in order to reduce the amplitude of the
magnetization fluctuations. Fig.\ref{E2} shows that $e(t)$ for
the predictor-corrector method increases linearly with time whereas
the decomposition methods both yield a constant value for $e(t)$ as shown in
Fig.\ref{E2}. Fig.\ref{Mz2} displays the magnetization conservation
for the predictor-corrector method $(\delta t = 0.01/J)$ and the
second order decomposition method according to Eq.(\ref{eA2BA2}) for
$\delta t = 0.04/J$. The predictor-corrector method conserves $m(t)$
{\em exactly}, whereas the second order decomposition causes
fluctuations of $m(t)$ on all time scales. The apparent deviation of
the temporal average of the magnetization from the exact value is due
to the finite sample time; for times greater than $2000/J$ the
deviation changes sign and fluctuations on time scales larger than
$800/J$ become visible. The temporal structure of
$m(t)$ for the decomposition methods given by Eq.(\ref{eA2BA2}) for
$\delta t = 0.04/J$ and Eq.(\ref{epABA}) for $\delta t = 0.2/J$ is
displayed in Fig.\ref{Mz4}. It is remarkable that the fourth order
decomposition gives a substantially better magnetization conservation
than the second order decomposition despite the very large time step.
In order to achieve the same overall accuracy of the magnetization
conservation with the second order decomposition a time step $\delta t
< 0.02/J$ is neccessary. A single integration of the equations of
motion using Eq.(\ref{eA2BA2}) (second order decomposition) is about
twice as fast as the predictor-corrector method used here. The
more complex fourth order decomposition (see Eq.(\ref{epABA})),
however, is about $2.5$ times slower than the predictor corrector
method. Taking the increase in time step by factors of $4$ and $20$,
respectively, into account, we find that both decomposition methods
yield an eightfold speedup of the integration of the equation of
motion. If the overall quality of the magnetization conservation is
also taken into account, there is a clear advantage for the fourth
order decomposition according to Eq.(\ref{epABA}) for the isotropic
case $D = 0$.

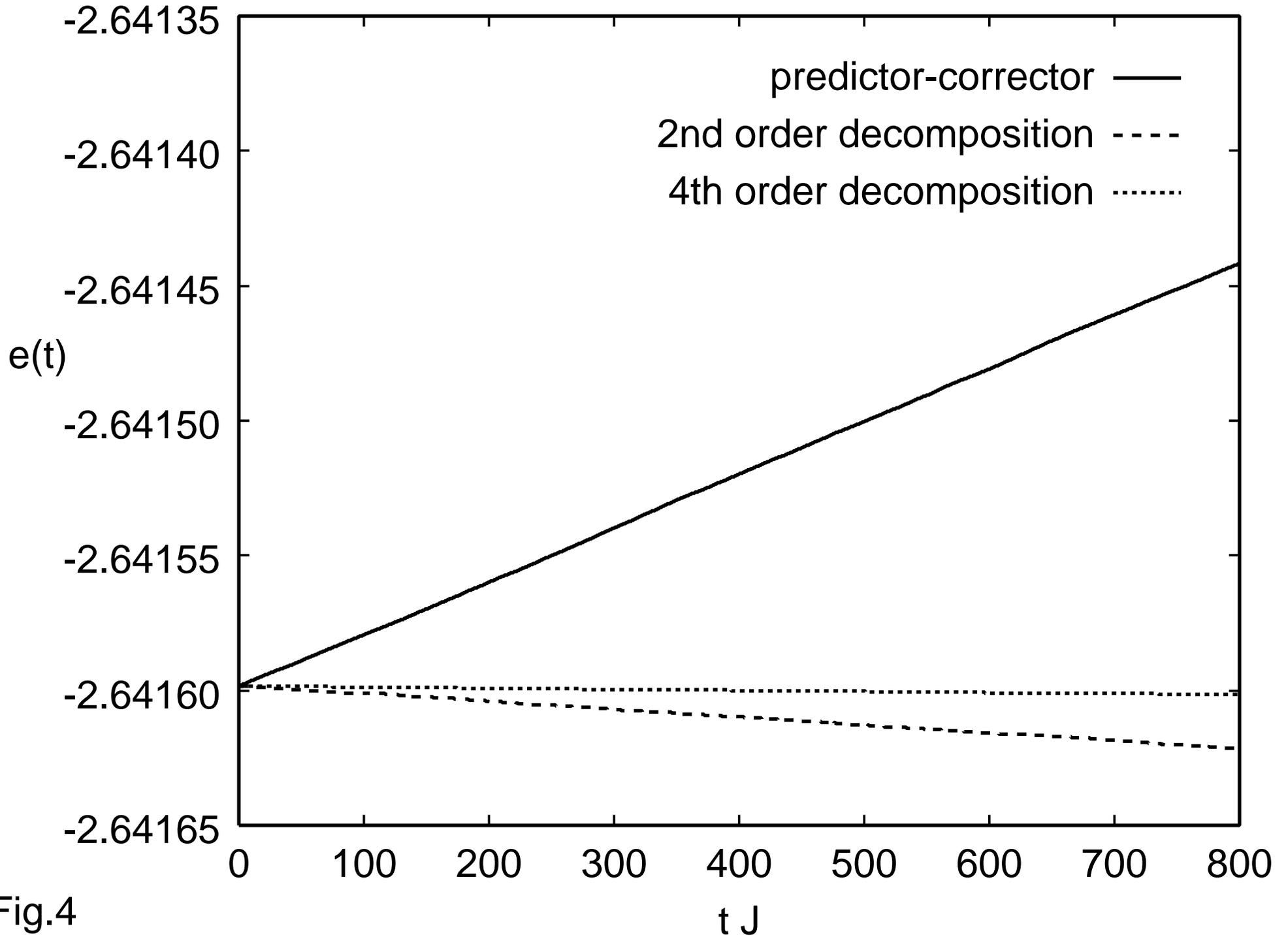
\begin{figure}[t]
\caption{
Energy $e(t)$ per spin for the predictor-corrector method (see
Eqs.(\protect\ref{AdBash}) and (\protect\ref{AdMoul})) for the time
step $\delta t = 0.01/J$ and $D = J$ (solid line). The decomposition
scheme given by Eq.(\protect\ref{eA2BA2}) for $\delta t = 0.04/J$
shows a decrease of $e(t)$ (dashed line) and
Eq.(\protect\ref{epABA})) for $\delta t = 0.2/J$ also yields a
decrease of $e(t)$ (dotted line). For these parameters the decomposition
methods show better energy conservation than the
predictor-corrector method.
\label{E2D}}
\end{figure}

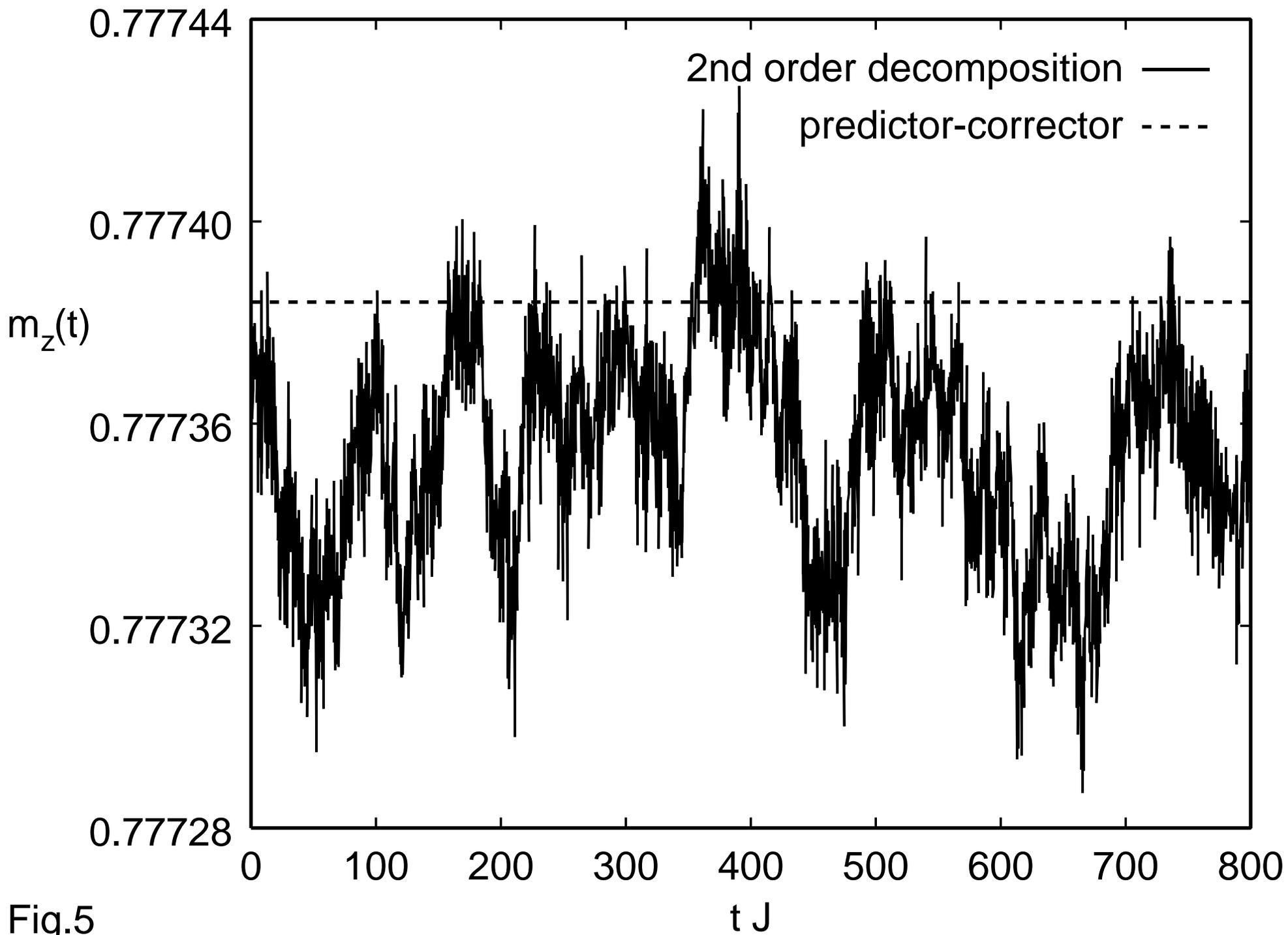
\begin{figure}[t]
\caption{
Magnetization $m_z(t)$ per spin for the predictor-corrector method (see
Eqs.(\protect\ref{AdBash}) and (\protect\ref{AdMoul})) for the time
step $\delta t = 0.01/J$ (dashed line) and the second order
decomposition scheme (see Eq.(\protect\ref{eA2BA2})) for the time step
$\delta t = 0.04/J$ (solid line) and $D = J$. The qualitative behavior
of $m_z(t)$ closely resembles the behavior in the isotropic case shown in
Fig.\protect\ref{Mz2}.
\label{Mz2D}}
\end{figure}

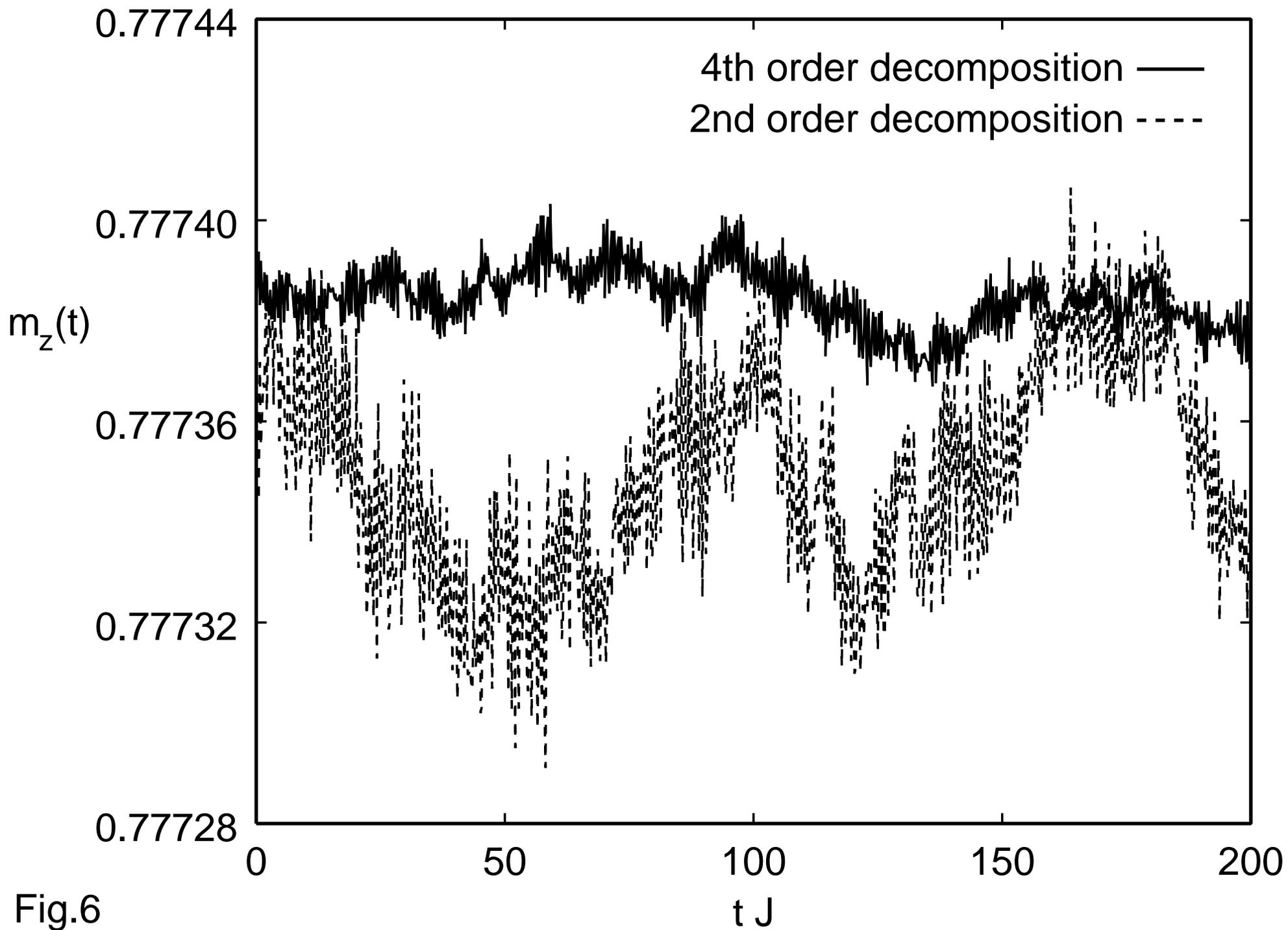
\begin{figure}[t]
\caption{
Magnetization $m_z(t)$ per spin for the fourth order decomposition
method (see Eq.(\protect\ref{epABA})) with $\delta t = 0.2/J$ (solid
line) and for the second order decomposition method (see
Eq.(\protect\ref{eA2BA2})) with $\delta t = 0.04/J$ (dashed line) for
$D = J$. The typical temporal structure of the fluctuations of
$m_z(t)$ is very similar to the isotropic case shown in
Fig.\protect\ref{Mz4}.
\label{Mz4D}}
\end{figure}

We now turn to the strongly anisotropic case $D = J$. The
predictor-corrector method can be applied as before, but the
decomposition scheme needs a major modification because the spin
rotation axis depends on the spin value $S_k^z$ at the future time $t
+ \delta t$ (see Eq.(\ref{OmegaD})). As already pointed out in Sec.III
this gives rise to a self consistency problem which can be
solved iteratively, where the quality of the energy conservation
depends on the number of iterations performed. For the second order
decomposition (see Eq.(\ref{eA2BA2})) and the time step $\delta t =
0.04/J$ two iterations are sufficient to obtain a better energy
conservation than the predictor-corrector method. This means
that a single integration of the equations of motion using the second
order decomposition takes twice as long as for $D = 0$, so that its
advantage in speed only comes from the increase in the time step,
which is still a factor of four. For the fourth order decomposition
according to Eq.(\ref{epABA}) with $\delta t = 0.2/J$ six iterations
are needed to obtain energy conservation to within six significant
digits, which makes one integration with the fourth order 
decomposition 15 times slower than one integration with the
predictor-corrector method. From the increase of $\delta t$ by a
factor 20 only a 30\% gain in speed remains for Eq.(\ref{epABA}).
The number of iterations needed decreases with $\delta t$, but this
decrease does not compensate the loss in speed due to the smaller time
step. However, one still obtains a greatly improved energy
conservation. The direct comparison of the energy conservation is
shown in Fig.\ref{E2D}. All three methods behave in a similar way, the
change in energy is basically linear with time. The reason for this is
the iterative nature of all three methods in the case $D \neq 0$. The
magnetization fluctuations shown in Fig.\ref{Mz2D} for the
predictor-corrector and the second order decomposition method $(\delta
t = 0.04/J)$ reveal the same behavior of both methods as shown in
Fig.\ref{Mz2} for $D = 0$. The direct comparison between the second
and fourth order decomposition for $\delta t = 0.04/J$ and $\delta t =
0.2/J$, respectively, is displayed in Fig.\ref{Mz4D}. The overall
accuracy of the magnetization conservation appears to be independent
of $D$ for both decomposition methods. If the emphasis is put on
overall energy conservation {\em and} speed, the second order
decomposition given by Eq.(\ref{eA2BA2}) has some advantages over the
predictor-corrector method. If the emphasis is put on energy
conservation alone, the fourth order decomposition given by
Eq.(\ref{epABA}) shows the best performance, but it is only slightly
faster than the predictor-corrector method.

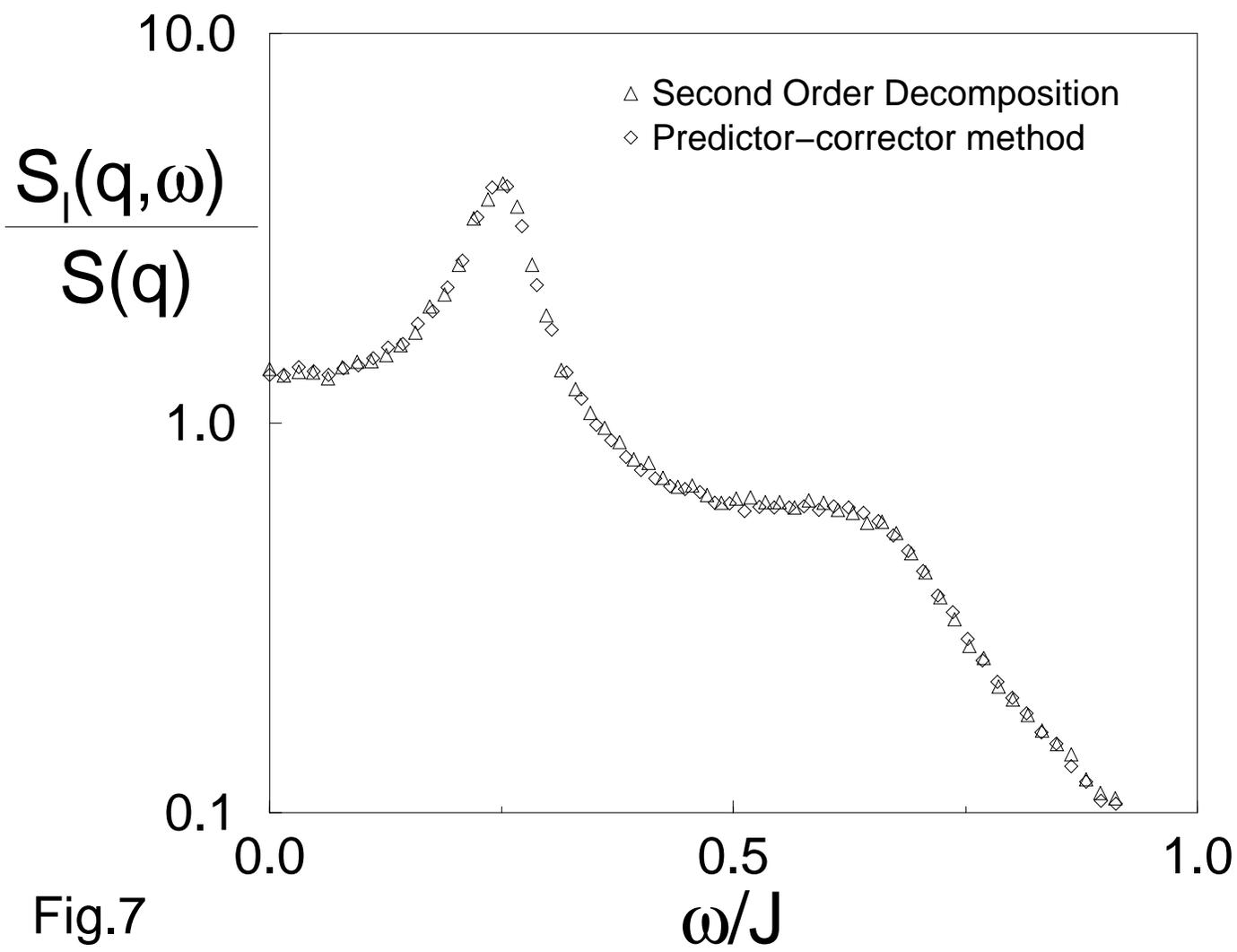
\begin{figure}[t]
\caption{
Longitudinal dynamic structure factor $S_l({\bf q},\omega)$ of an
isotropic Heisenberg ferromagnet for $T = 0.8 T_c$ and $|{\bf q}| =
\pi/5$ in the $(100)$ direction on a simple cubic lattice ($L = 10$,
periodic boundary conditions) normalized to the static structure
factor $S_l({\bf q}) \equiv \int_0^{\infty} S_l({\bf q},\omega) d \omega$
for the predictor-corrector method (diamonds) and the second order
decomposition method (triangles) for time steps $\delta t = 0.01/J$
and $\delta t = 0.04/J$, respectively (see main text). The statistical
error (one standard deviation) is smaller than the symbol size.
\label{Sql}}
\end{figure}

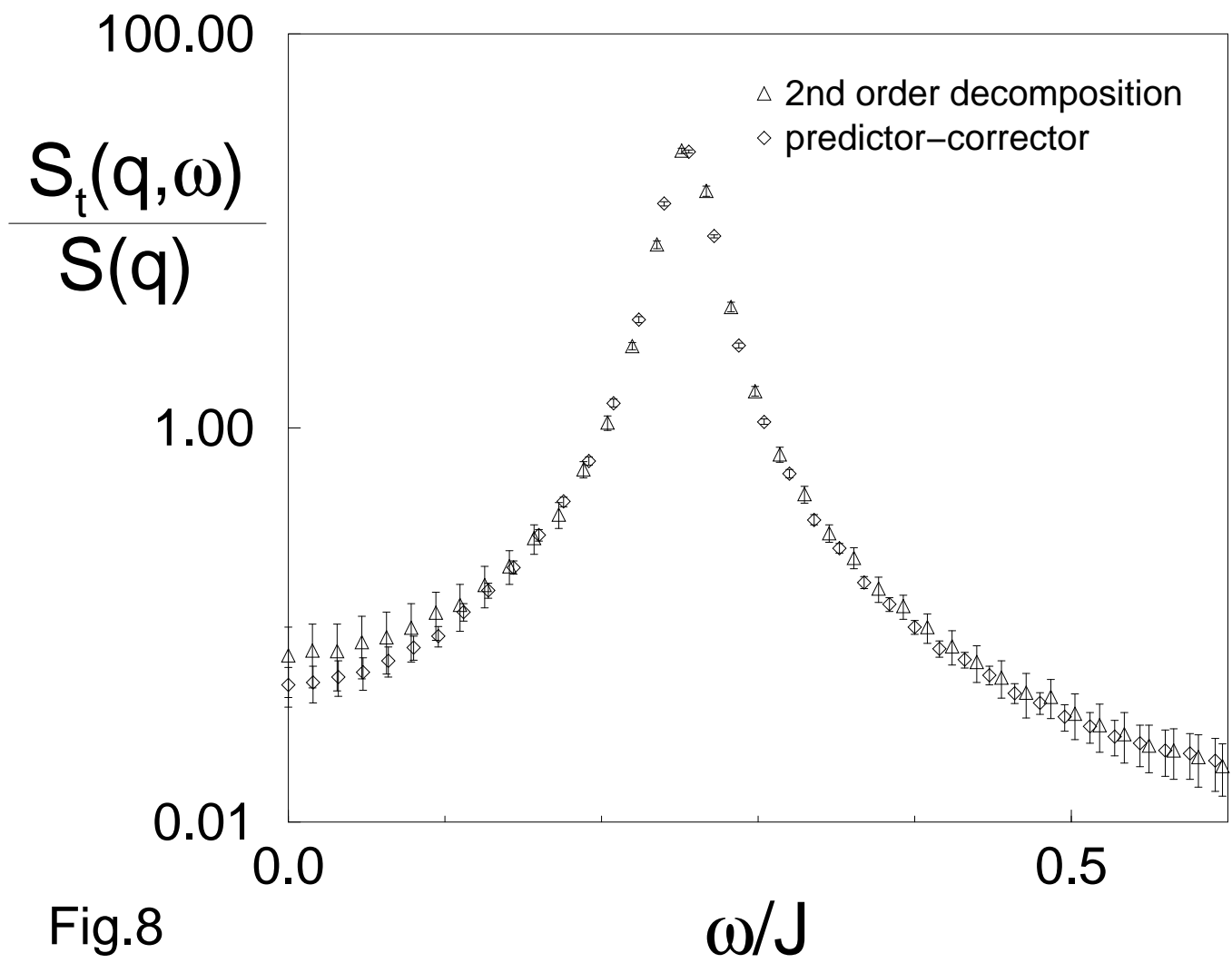
\begin{figure}[t]
\caption{
Transverse dynamic structure factor $S_t({\bf q},\omega)$ of an
isotropic Heisenberg ferromagnet for $T = 0.8 T_c$ and $|{\bf q}| =
\pi/5$ in the $(100)$ direction on a simple cubic lattice ($L = 10$,
periodic boundary conditions) normalized to the static structure
factor $S_t({\bf q}) \equiv \int_0^{\infty} S_t({\bf q},\omega) d \omega$
for the predictor-corrector method (diamonds) and the second order
decomposition method (triangles) for time steps $\delta t = 0.01/J$
and $\delta t = 0.04/J$, respectively (see main text). The error bars
represent one standard deviation.
\label{Sqt}}
\end{figure}

We close this section with a direct comparison of the longitudinal
($S_l({\bf q},\omega)$) and transverse ($S_t({\bf q},\omega)$) components 
of the dynamic structure factor for the isotropic Heisenberg
ferromagnet (see Eq.(\ref{H}) for $D = 0$) on a simple cubic lattice
($L = 10$, periodic boundary conditions) at the temperature $T = 0.8
T_c$ in the $(100)$ direction. In order to measure $S_l({\bf q},\omega)$
one has to consider the projections of all spins onto the direction of
the magnetization ${\bf M}$ of the initial configuration. Note that in
general ${\bf M}$ for a given initial (equilibrium) spin configuration
will be nonzero even in a finite system. Only the {\em average} $\langle
{\bf M} \rangle$ of the magnetization over sufficiently many equilibrium
configurations yields $\langle {\bf M} \rangle = 0$ within the
statistical errors. The spin components transverse to ${\bf M}$ then give
access to $S_t({\bf q},\omega)$. The results are shown in Figs.\ref{Sql}
and \ref{Sqt} for ${\bf q} = (\pi/5,0,0)$, where $S_l({\bf q},\omega)$
and $S_t({\bf q},\omega)$ have been normalized to the static structure
factor $S_{l,t}({\bf q}) \equiv \int_0^{\infty} S_{l,t}({\bf
q},\omega) d \omega$. The diamonds represent the result for the
predictor-corrector method $(\delta t = 0.01/J)$ and the triangles show
the result for the second order decomposition method (see
Eq.(\ref{eA2BA2}), $\delta t = 0.04/J$). For both methods the equations
of motion have been integrated to $800/J$ and averages have been taken
over 1000 initial configurations, where the time displaced correlation
functions have been measured to $400/J$. The statistical error
indicated by the error bar represents one standard deviation. For the
longitudinal structure factor shown in Fig.\ref{Sql} the statistical
error is smaller than the symbol size. The overall agreement of the
data is very good. The main maximum corresponds to the spin wave peak
and is located at $\omega_0 = 0.25 J$ in all cases. The frequency
resolution is $\delta \omega = 2\pi J / 400 \simeq 0.016 J$. The
shoulder like feature at $\omega \simeq 0.5 J$ in $S_l({\bf
q},\omega)$ (see Fig.\ref{Sql}) is due to many-spin wave processes, the
description of which is beyond the scope of this article. Both methods
have spent about the same CPU time on generating the initial
configurations, but with the time steps given above, the integration
of the equations of motion with the second order decomposition method
is about eight times faster than with the predictor-corrector method.
The discrepancy between the errorbars shown in Fig.\ref{Sqt} is of
statistical origin and indicates the statistical uncertainty of the
errorbars themselves. Note, however, that the discrepancy only occurs
in a frequency region well outside the spin wave peak, where the signal
level is almost three orders of magnitude below its peak value.

\section{Summary}
Starting from a well developed and established predictor-corrector
method for the numerical integration of the equations of motion of a
classical spin system as a timing and precision standard we devised
and tested an alternative class of algorithms which is based on
Suzuki-Trotter decompositions of exponential operators. Our findings
can be summarized as follows.
\begin{description}
\item{1. Advantages of the predictor-corrector method}

The greatest advantage of the predictor-corrector method is its
versatility and its capability to conserve the magnetization exactly.
Isotropic and anisotropic spins systems with nearest and next-nearest
neighbor interactions can be treated within one and the same numerical
approach. The decomposition of the lattice into sublattices, which is
the basis for the decomposition method, depends on the range of the
interactions, so that this numerical approach is far less general than
the predictor-corrector method. Crystal field anisotropies leave the
performance of the predictor-corrector method almost unaffected,
whereas the decomposition method suffers from a drastic reduction in
speed.

\item{2. Advantages of the decomposition method}

The greatest advantage of the decomposition method is its capability
for handling large time steps and the exact conservation of spin
length. In the absence of anisotropies it also conserves
the energy exactly and it maintains reversibility. For anisotropic
Hamiltonians energy conservation and reversibility can be obtained to
a high accuracy using iterative schemes; exact magnetization
conservation, however, is lost. The time steps typically used
are unaccessible by the predictor-corrector method due to the
lack of exact energy conservation within the algorithm. The resulting
speedup of a spin dynamics simulation gives access to larger lattices,
provided a second order decomposition is sufficiently accurate for
the problem under investigation. In simple cases the fourth order
decomposition yields very accurate results even for time steps
an order of magnitude larger than typical time steps used
for the predictor-corrector method.
\end{description}

A general recommendation for one or the other method cannot be given.
If, however, special interactions are the main objective of the
investigation, one should resort to the predictor-corrector method
because of its generality. If the interactions in the system are
standard two-spin exchange interactions and the objective is to study
large systems, one should consider the decomposition method because of
its speed and its built-in energy and spin length conservation.

\acknowledgments
M. Krech gratefully acknowledges financial support of this work
through the Heisenberg program of the Deutsche Forschungsgemeinschaft.
This research was supported in part by NSF grant \#DMR - 9405018
and the Pittsburgh Supercomputer Center.
\bigskip

\noindent
{\bf Note added in proof}

While this work was in press we have learned that a method equivalent to the
one presented here had been developed independently by Frank, Huang, and
Leimkuhler (J. Frank, W. Huang, and B. Leimkuhler, Journal of Computational
Physics {\bf 133}, 160 (1997).) Frank, Huang, and Leimkuhler denote their
method as {\em Staggered Red-Black Method} and employ the same decomposition
technique which is described in this work.

\end{document}